# Spatially resolved optical absorption spectroscopy of single- and few-layer MoS$_2$ by hyperspectral imaging

Andres Castellanos-Gomez[1,*], Jorge Quereda[2], Herko P. van der Meulen[3], Nicolás Agraït[1,2], and Gabino Rubio-Bollinger[2,4,*]

[1]Instituto Madrileño de Estudios Avanzados en Nanociencia (IMDEA-Nanociencia), 28049 Madrid, Spain.

[2]Dpto. de Física de la Materia Condensada, Universidad Autónoma de Madrid, 28049 Madrid, Spain.

[3]Dpto. de Física de Materiales, Universidad Autónoma de Madrid, 28049 Madrid, Spain.

[4]Condensed Matter Physics Center (IFIMAC), Universidad Autónoma de Madrid, E-28049 Madrid, Spain.

[*] E-mail: andres.castellanos@imdea.org , gabino.rubio@uam.es

**Abstract**

The possibility of spatially resolving the optical properties of atomically thin materials is especially appealing as they can be modulated at the micro- and nanoscale by reducing their thickness, changing the doping level or applying a mechanical deformation. Therefore, optical spectroscopy techniques with high spatial resolution are necessary to get a deeper insight into the properties of two-dimensional materials. Here we study the optical absorption of single- and few-layer molybdenum disulfide (MoS$_2$) in the spectral range from 1.24 eV to 3.22 eV (385 nm to 1000 nm) by developing a hyperspectral imaging technique that allows one to probe the optical properties with diffraction limited spatial resolution. We find hyperspectral imaging very suited to study indirect bandgap semiconductors, unlike photoluminescence that only provides high luminescence yield for direct gap semiconductors. Moreover, this work opens the door to study the spatial variation of the optical properties of other two-dimensional systems, including non-semiconducting materials where scanning photoluminescence cannot be employed.

## 1. Introduction

Atomically thin semiconductors hold the promise to complement graphene in those applications where graphene's lack of a band gap hampers its use. Single-layer MoS$_2$, the most studied 2D semiconductor to date,[1–4] has already being applied in nanoelectronic devices like field-effect transistors,[5] non-volatile memories [6–8] or logic circuits.[9,10] The direct gap of single-layer $MoS_2$, in the visible range of the electromagnetic spectrum,[11,12] has also triggered the interest on its optical properties.

In the last years, scanning confocal microscopy based optical spectroscopy techniques (especially micro-photoluminescence) has made it possible to observe interesting optical phenomena such as quantum confinement induced direct-to-indirect band gap transition, [11–13] photoluminescence emission from charged excitons [14–16] and valley-polarized photoluminescence yield.[16–20] Photoluminescence spectroscopy, however, is a very inefficient technique to study the optical properties of indirect band gap semiconductors (like few-layer $MoS_2$) and thus transmittance or absorption spectroscopy based techniques would be preferred instead. Nonetheless, spatially resolved absorption spectroscopy techniques are still very scarce and are mainly based on non-trivial modifications of expensive scanning confocal microscopy setups.[21]

In this work we probe the local optical absorption of single- and few-layer molybdenum disulfide ($MoS_2$) by hyperspectral imaging. This technique allows one to study the optical properties of $MoS_2$ with diffraction limited spatial resolution in a very important range of the electromagnetic spectrum ranging from near-ultraviolet (NUV) to near-infrared (NIR) with interest for optoelectronics, night-vision imaging and photovoltaics. We find that this technique is especially convenient to study the optical properties of multilayer $MoS_2$ and other indirect bandgap 2D materials that typically yield weak luminescence signals and thus are challenging to study with photoluminescence.

## 2. Experimental section

Atomically thin $MoS_2$ samples have been prepared by mechanical exfoliation of bulk $MoS_2$ (*SPI Supplies*, *429ML-AB*) with Nitto tape (*Nitto Denko Co., SPV224 clear*).

Poly(dimethylsiloxane) (PDMS) has been chosen as substrate because of its transparency and its low interaction with $MoS_2$. In fact, it has been recently shown that the optical properties of $MoS_2$ on PDMS resembles those of free-standing $MoS_2$ because of the reduced charge transfer between the PDMS and $MoS_2$, much smaller than that of $MoS_2$ on $SiO_2$.[22]

In order to study the local optical absorption of the fabricated $MoS_2$ flakes, we have developed a hyperspectral imaging setup by modifying an optical microscope (*Nikon Eclipse LV-100*), conventionally used in many laboratories to identify 2D materials. A fiber bundle is attached through the sample stage so transparent samples can be directly placed on top the output of the bundle to be studied in transmission mode using the light emitted from the fibers. The bundle is connected to a tunable monochromatic light source (a combination of a tungsten lamp and a monochromator) which allows one to select the excitation wavelength with a bandwidth down to 1-2 nm. A monochrome CMOS camera (*Edmund Optics, EO-5012 Monochrome USB 3.0 Camera*) is attached to the trinocular of the microscope for detection. Note that this camera model does not incorporate the IR blocking filter (present in most CMOS cameras) making it possible to detect light with wavelength above 700 nm. A spectrum of the fiber bundle output light, measured with the monochrome camera through the microscope optics while the excitation wavelength is swept, is presented in the Supporting Information (Figure S1) and it can be employed to determine the detectable photon energy range of the experimental setup: 1.2 eV to 3.3 eV (or 375 nm to 1030 nm). Figure 1a shows a cartoon of the experimental setup used for the hyperspectral imaging. Note that this setup can be implemented by relatively inexpensive (<15000$) modifications of a conventional optical microscope.

**3. Results and discussion**

The hyperspectral imaging is carried out by sweeping the excitation wavelength in steps and acquiring a transmission mode image for each wavelength. The collected data is then arranged in a three-dimensional matrix, being the first two matrix indexes the *X* and *Y* spatial coordinates and the third index (*λ*) the wavelength (see the sketch in Figure 1b). Spectral information of a certain sample region can be directly obtained by plotting all the elements along the wavelength dimension *λ* for given *X* and *Y* coordinates. This corresponds to a vertical cut along the wavelength axis in the matrix sketched in Figure 1b. Figure 1c shows a cartoon with two spectra extracted from two regions in the sample: bare substrate (red) and $MoS_2$ (blue). Quantitative transmittance (*T*) information of the $MoS_2$ region can be obtained by dividing both spectra: $T = I_{flake}/I_{subs}$ (see the inset in Figure 1c), where $I_{flake}$ is the intensity acquired at a MoS2 region and $I_{subs}$ the intensity measured onto the bare substrate. The absorbance *A* can be obtained from the transmittance *T* as: $A = -\log_{10}(T)$.

Figure 2 shows a sequence of transmission mode optical images of a $MoS_2$ flake acquired at different excitation wavelengths, selected with the tunable monochromatic light source. Although Figure 2 only shows six images, a total number of 123 images were collected from 385 nm (3.22 eV) to 1000 nm (1.24 eV) in steps of 5 nm. The whole data acquisition takes less than 30 minutes. Regions with different number of layers, as determined by Raman spectroscopy and photoluminescence (see Figure S2 and Figure S3 in the Supporting Information), have been highlighted in the first panel of Figure 2. At illumination wavelength longer than 670 nm the absorption of the monolayer region decreases abruptly. This observation agrees with the fact that the bandgap of single-layer $MoS_2$ is 1.85 eV (corresponding to a cut-off wavelength of 670 nm) and it monotonically decreases while increasing the number of $MoS_2$ layers until reaching the bulk band gap value of ~1.35 eV.

In order to obtain quantitative spectral information, the sequence of 123 images is arranged as previously discussed for Figure 1b. Then, intensity *vs.* wavelength spectra are extracted at certain positions on the sample (similarly to the cartoon example in Figure 1c). All the spectra are normalized to an intensity *vs.* wavelength spectrum, obtained from a bare substrate region close to the MoS$_2$ region of interest, to determine the optical transmittance $T(\lambda) = I_{\text{flake}}(\lambda) / I_{\text{subs}}(\lambda)$. The optical absorbance of MoS$_2$ flakes 1L to 6L thick, calculated from the transmittance values, as a function of the excitation wavelength are displayed in Figure 3. The absorbance spectra have two prominent narrow peaks occurring at wavelengths ~605 nm and ~660 nm that correspond to the absorption due to the direct transitions at the K point of the Brillouin zone, associated to the generation of the B and A excitons respectively.[11,12] While the position of the A exciton peak wavelength monotonically red-shifts with the number of layers, the B exciton peak wavelength remains almost unaltered (see the inset in Figure 3). This observation is in agreement with scanning photoluminescence studies.[22] The spectra also show a broad peak around 440 nm. This feature is typically not observed in photoluminescence experiments which mostly use an excitation wavelength of ~500 nm. Recent reflectance and photocurrent spectroscopy experiments, however, present this feature (referred to as C-exciton peak) whose origin is still a subject of debate.[21,23] Interestingly, the position of this feature strongly depends on the number of layers. A detailed study of the thickness dependence of atomically thin MoS$_2$ will be reported somewhere else.

The fact that hyperspectral imaging provides one the optical transmittance/absorbance spectrum at every position of the flake can be exploited to generate maps where the spatial variation of a certain spectroscopic feature is probed. Figure 4 shows two maps displaying the spatial variation of the C and A exciton peak wavelength in the MoS$_2$ sample shown in Figure 2. By comparing Figure 4a and 4b with Figure 2 one can find a correspondence between the C and A exciton wavelength and the number of MoS$_2$ layers, as discussed from the spectra acquired for different number of layers (see Figure 3).

A linecut of the exciton peak wavelength across a step-edge, separating two regions with different number of layers, can be employed to estimate the spatial resolution. Below Figure 4a and 4b two line profiles measured along the step-edges highlighted with a white rectangle are shown. The experimental line profiles can be reproduced by a step function with a Gaussian broadening of 300 nm (for the C-exciton) and 420 nm (for the A-exciton). These values are in good agreement with the values expected for a system with diffraction limited spatial resolution, $d = \lambda/2NA$, where $d$ is the smallest resolvable feature $\lambda$ the illumination wavelength ($\lambda \sim 440$ nm for the C-exciton and $\lambda \sim 660$ nm for the A-exciton) and NA the microscope objective numerical aperture (NA = 0.8 in our experimental setup).

## 4. Conclusion

In summary, we have locally probed the optical absorption of single- and multilayer $MoS_2$ in an important range of the electromagnetic spectrum, spanning from 1.2 eV (NIR) to 3.3 eV (NUV). By developing a hyperspectral imaging technique, we are able of studying the optical properties of these 2D semiconductors with diffraction limited spatial resolution. We demonstrate this high spatial resolution by mapping the excitonic features, present in the absorbance spectra, at different regions of the $MoS_2$ sample. The hyperspectral imaging technique presented here can be applied to study a broad family of two-dimensional materials and its high spatial resolution opens the door to study the effect on the optical properties of localized strain,[24,25] inhomogeneous chemical composition [26–28] or doping level [29] in 2D systems.

**Acknowledgements**

This work was supported by the European Union (FP7) through the FP7-Marie Curie Project PIEF-GA-2011-300802 (STRENGTHNANO), by the Fundacion BBVA through the fellowship "I Convocatoria de Ayudas Fundacion BBVA a Investigadores, Innovadores y Creadores Culturales" ("Semiconductores ultradelgados: hacia la optoelectronica flexible"), by the Ministerio de Economia y Competitividad (MINECO) through the projects MAT2011-22997, MAT2014-53119-C2-1-R, MAT2011-25046 and MAT2014-57915-R and by the Comunidad Autónoma de Madrid through the project MAD2D-CM (S2013/MIT-3007).

FIGURES

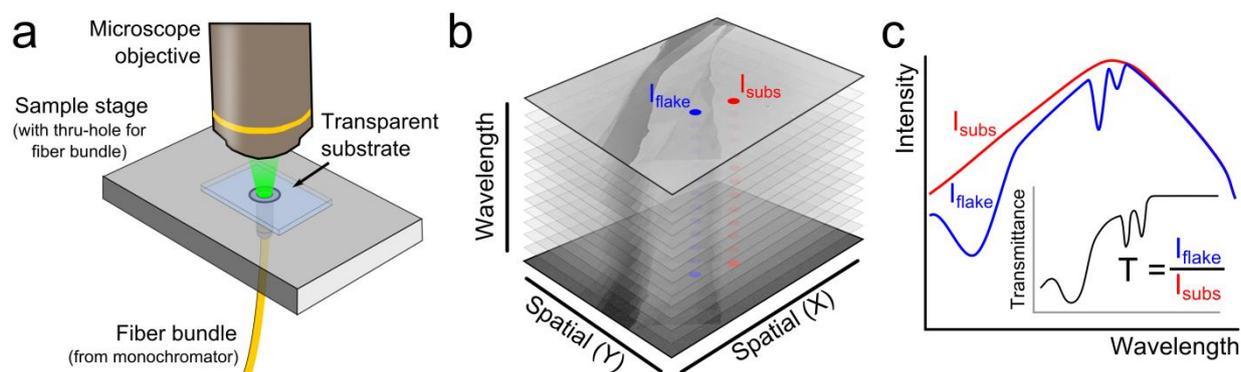

**Figure 1**. **Hyperspectral imaging of 2D materials.** (a) Sketch of the experimental configuration employed to carry out hyperspectral imaging in transmission mode, the sample is placed onto a fiber bundle connected to a tunable light source to select the illumination wavelength. A conventional optical microscope supplemented with a monochrome CMOS camera is used to acquire the transmission mode optical images at different illumination wavelengths. (b) Cartoon illustrating the data acquisition process. Several transmission mode optical microscopy images of a sample region are acquired while the illumination wavelength is varied. The resulting dataset can be arranged in a three-dimensional matrix. (c) By inspecting the intensity data along the wavelength ($\lambda$) dimension of the matrix displayed in (b), one can extract spectral information at a specific location on the sample. For example, (c) shows the spectral information acquired on a region of a $MoS_2$ flake and on a bare substrate region. The inset shows a cartoon of a transmittance spectrum obtained by normalizing the spectrum obtained on the $MoS_2$ flake region to the bare substrate spectrum.

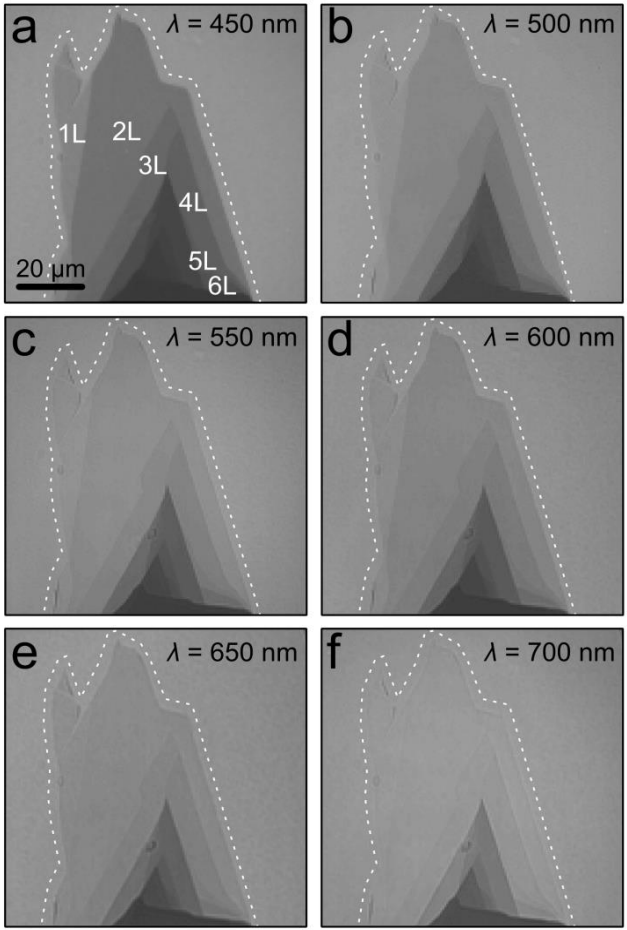

**Figure 2. Wavelength dependent transmission mode images of MoS$_2$.** Transmission mode optical microscopy images of a mechanically exfoliated MoS$_2$ flake acquired at different excitation wavelengths, selected by the tunable monochromatic light source. Although the whole dataset comprises 123 images with illumination wavelength spanning from 375 nm to 1000 nm in 5 nm steps, this Figure features six selected wavelengths. The first panel displays labels indicating the number of MoS$_2$ layers of the different regions of the flake. The dotted white line highlights the border of the MoS$_2$ flake to facilitate the comparison between images acquired at different wavelengths.

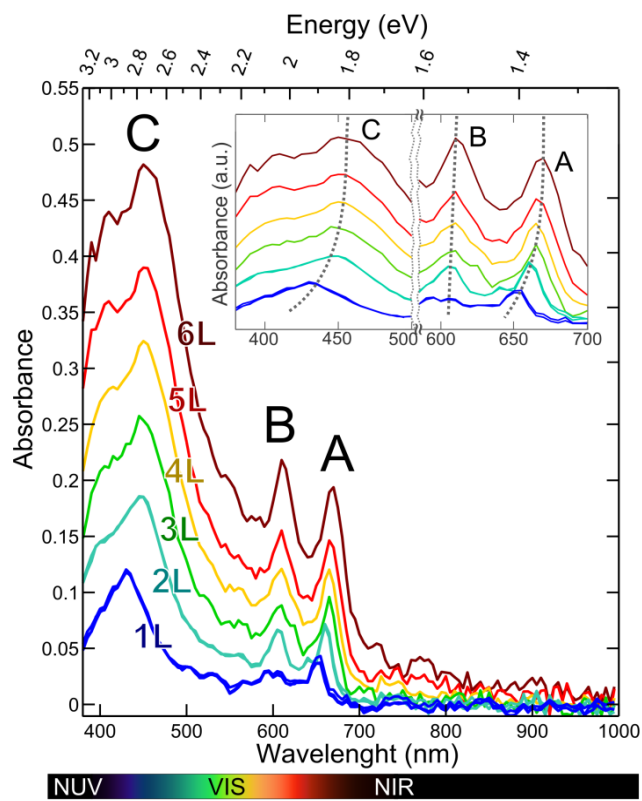

**Figure 3. Absorbance spectra of single- and few-layer MoS$_2$.** Absorbance *vs.* wavelength spectra, extracted from the sequence of images of a MoS$_2$ flake with regions of different thicknesses (ranging from single-layer to six-layers), acquired at different illumination wavelengths. The labels A, B and C indicate the features attributed to the generation of the A, B and C excitons. The inset shows a detail of the thickness dependent wavelength of the excitonic peaks.

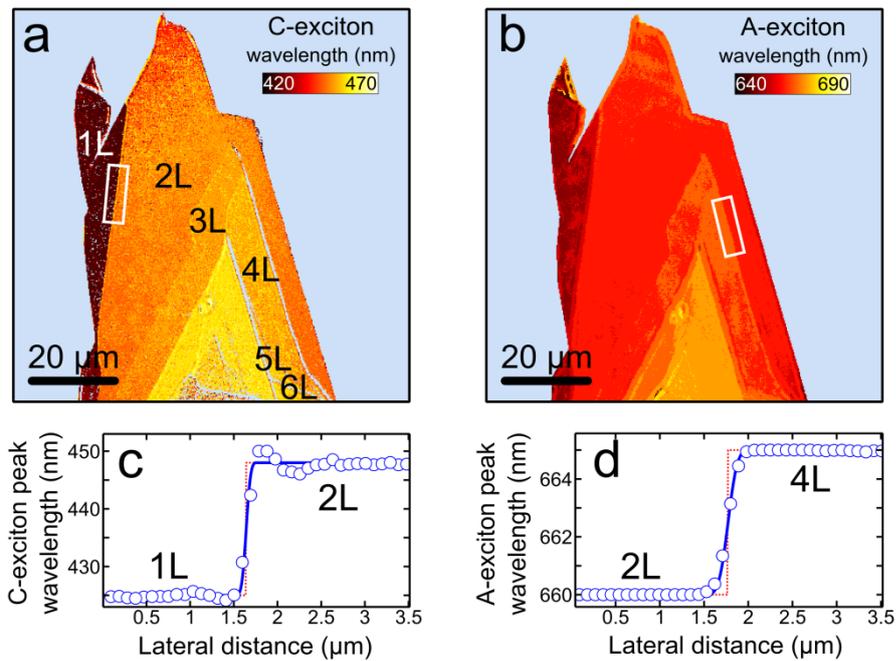

**Figure 4. Spatially resolved optical properties of MoS$_2$.** (a) and (b) false color map displaying the position of the C and A exciton peaks, respectively, at the different regions of the MoS$_2$ flake showed in Figure 2. (c) and (d) show a lineprofile acquired at the interface between two regions with different number of layers, indicated with a white rectangle in (a) and (b). From the spatial variation of the exciton peak position in (c) and (d) one can estimate the spatial resolution: 300 nm for the C-exciton and 420 nm for the A-exciton (see the text for more details).

# Supporting information

# Spatially resolved optical absorption spectroscopy of single- and few-layer MoS$_2$ by hyperspectral imaging

*Andres Castellanos-Gomez[1,*], Jorge Quereda[2], Herko P. van der Meulen[3], Nicolás Agraït[1,2], and Gabino Rubio-Bollinger[2,4,*].*

[1]Instituto Madrileño de Estudios Avanzados en Nanociencia (IMDEA-Nanociencia), 28049 Madrid, Spain.
[2]Dpto. de Física de la Materia Condensada, Universidad Autónoma de Madrid, 28049 Madrid, Spain.
[3]Dpto. de Física de Materiales, Universidad Autónoma de Madrid, 28049 Madrid, Spain.
[4]Condensed Matter Physics Center (IFIMAC), Universidad Autónoma de Madrid, E-28049 Madrid, Spain.

Table of contents:

1) Experimental setup wavelength range
2) Raman spectroscopy of MoS$_2$ flakes
3) Photoluminescence of MoS$_2$ flakes
4) Comparison between the spectra extracted from hyperspectral and multispectral imaging methods

*Experimental setup bandwidth:*

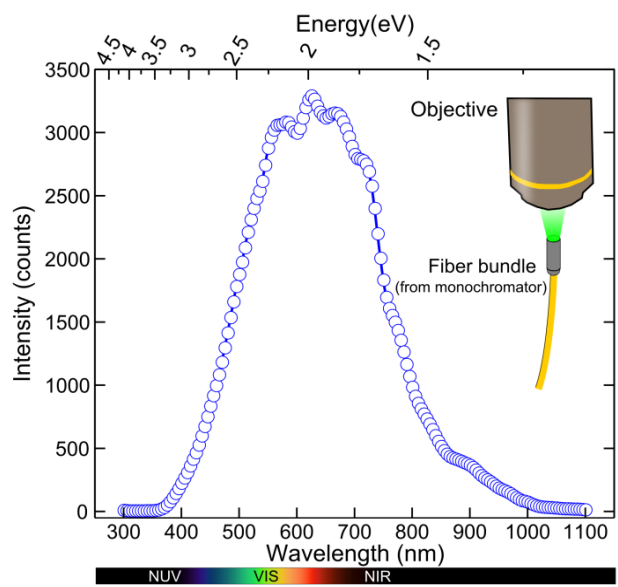

**Figure S1. Measurement of the hyperspectral setup wavelength range.** The wavelength range of the experimental setup has been determined by directly acquiring images of the while the illumination wavelength is swept with the tunable monochromatic light source. The Figure shows the intensity recorded at the center of an individual fiber at each illumination wavelength.

*Raman spectroscopy of MoS$_2$ flakes*

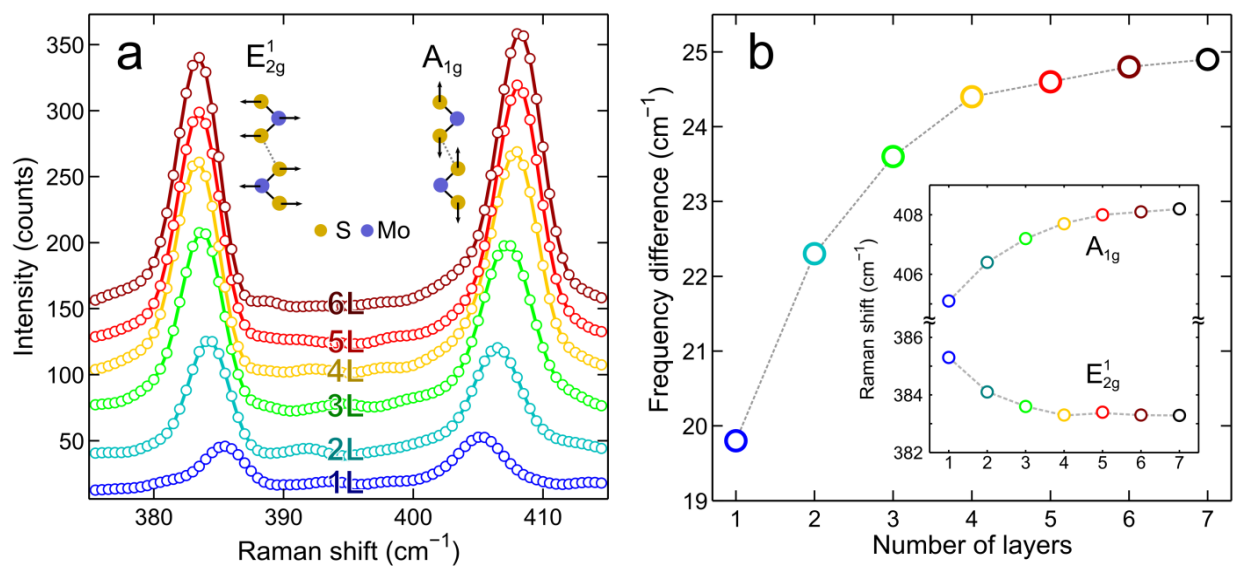

**Figure S2. Raman spectra of exfoliated MoS$_2$ flakes on the PDMS substrate.** (a) Raman spectra measured for MoS$_2$ flakes deposited onto PDMS substrates with different number of layers. (b) Frequency difference between the $E^1_{2g}$ and the $A_{1g}$ modes as a function of the number of layers. This value increases monotonically with the number of layers from (~19.8 cm$^{-1}$ for single layers to ~ 25 cm$^{-1}$ for bulk MoS$_2$).

*Photoluminescence of MoS₂ flakes*

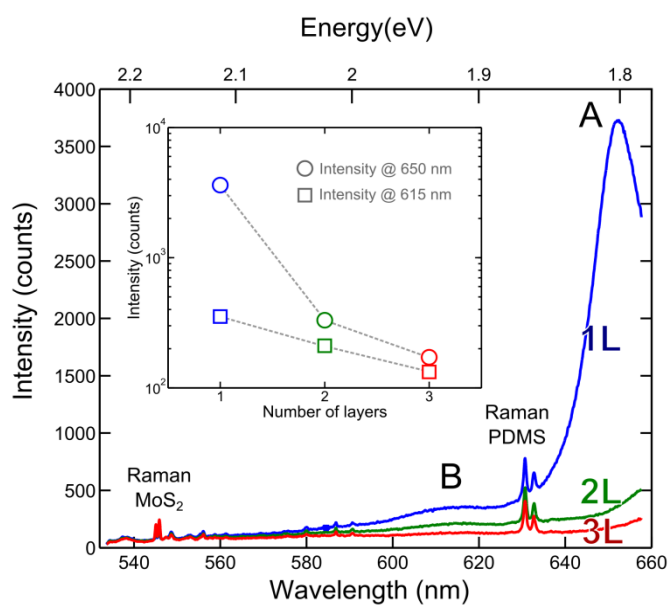

**Figure S3. Photoluminescence characterization of the MoS$_2$ flakes on the PDMS substrate.** Photoluminescence spectra acquired for single-layer, bilayer and trilayer MoS$_2$ flakes deposited onto a PDMS substrate. The photoluminescence yield corresponding to the B and A excitons are visible and their intensity is strongly thickness dependent. The inset shows the intensity of the photoluminescence spectra at two wavelengths (615 nm and 650 nm) as a function of the number of layers. Monolayer MoS$_2$ can be readily distinguished from bilayer and trilayer flakes by its strong luminescence yield.

*Comparison between the spectra extracted from hyperspectral and multispectral imaging methods*

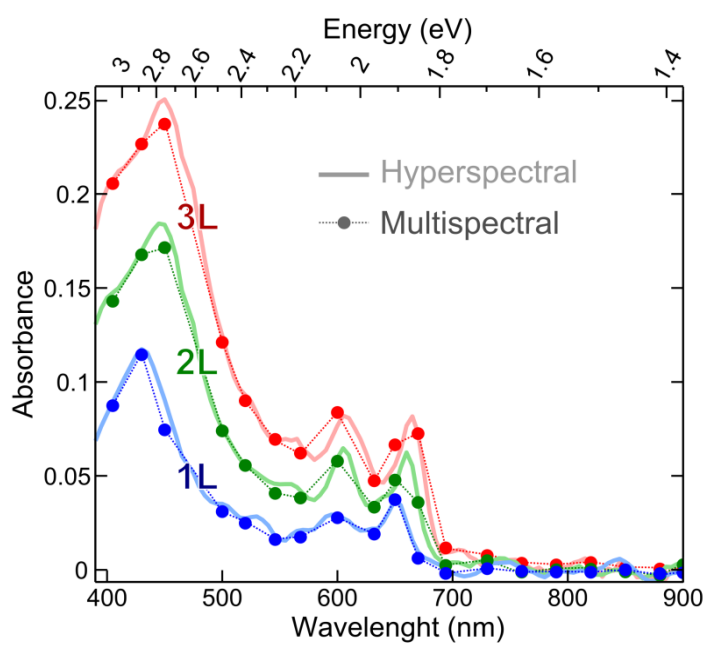

**Figure S4. Comparison between the hyperspectral and multispectral imaging.** Alternatively to the use of the tunable monochromatic light source, one can also select the illumination wavelength with narrow bandwidth filters. This Figure illustrates the results obtained by imaging single-layer, bilayer and trilayer MoS$_2$ with 19 bandpass filters (10 nm of FWHM). Although the agreement is excellent between these two ways of illuminating the sample, the use of bandpass filters hampers the spectral resolution of the measurements and fine details can only be captured with a much finer wavelength sweep (as that provided by the hyperspectral imaging setup developed here).